# Observation of Single-Molecule Raman Spectroscopy Enabled by Synergic Electromagnetic and Chemical Enhancement

Haiyao Yang, Haoran Mo, Jianzhi Zhang, Lihong Hong, and Zhiyuan Li✉

There has been a long fundamental pursuit to enhance and levitate the Raman scattering signal intensity of molecule by a huge number of ~14-15 orders of magnitude, to the level comparable with the molecule fluorescence intensity and truly entering the regime of single-molecule Raman spectroscopy. In this work we report unambiguous observation of single-molecule Raman spectroscopy via synergic action of electromagnetic and chemical enhancement for rhodamine B (RhB) molecule absorbed within the plasmonic nanogap formed by gold nanoparticle sitting on the two-dimensional (2D) monolayer $WS_2$ and 2 nm $SiO_2$ coated gold thin film. Raman spectroscopy down to an extremely dilute value of $10^{-18}$ mol/L can still be clearly visible, and the statistical enhancement factor could reach 16 orders of magnitude compared with the reference detection sample of silicon plate. The electromagnetic enhancement comes from local surface plasmon resonance induced at the nanogap, which could reach ~10-11 orders of magnitude, while the chemical enhancement comes from monolayer $WS_2$ 2D material, which could reach 4-5 orders of magnitudes. This synergic route of Raman enhancement devices could open up a new frontier of single molecule science, allowing detection, identification, and monitor of single molecules and their spatial-temporal evolution under various internal and external stimuli.

## Introduction

Raman spectroscopy is a traditional and powerful means to observe vibrational, rotational, and other low-frequency modes in materials via inelastic scattering against light, and enable molecule and material analysis and identification [1-7]. After 90 years of development, it still attracts tremendous interest to probe single-molecule (SM) Raman spectroscopy, which is not just an ultrasensitive version of traditional Raman spectroscopy, but rather opens up a new window to observe subtle spectroscopic phenomena in single molecules without statistical average, enabling much more clarified observation of molecule science details [8-11].

The enhancement factor (EF) of Raman spectra is the ratio of Raman signal with and without substrates in the same experimental environment, normalized for the number of molecules probed:

$$EF = \frac{I_E/N_{surf}}{I_R/N_{vol}} \qquad (1)$$

where $I_R$ and $I_E$ are the Raman intensity for $N_{vol}$ molecules and enhanced Raman intensity for $N_{surf}$ molecules, respectively. Since Nie, Kneipp and their co-workers claimed SM Raman detection for the first time in 1997 [11,12], and Xu's team reported a $10^{10}$ EF for single hemoglobin molecule in 1999 [13], all through the scheme of surface-enhanced Raman spectroscopy (SERS) mediated by surface plasmon polariton (SPP) excitation and action, multiple researches paving towards the road of SM-SERS have been accomplished [14-24], among these, the highest EF is $10^{12}$ achieved by Li et al. [21].

It is generally acknowledged that Raman signal needs a huge EF of ~14-15 orders of magnitude to maintain a signal intensity comparable with molecule fluorescence and truly enter the regime of single-molecule detection sensitivity. This criterion has set a huge obstacle to frustrate and obstruct most contemporary macroscopic, mesoscopic, and microscopic strategies of Raman scattering enhancement [7], including two most prestigious schemes, the electromagnetic enhancement (EME) and chemical enhancement (CME) [25-27]. EME is mainly induced by mesoscopic local surface plasmon resonance (LSPR) effect in metallic nanoparticles and nanostructures with their free electron gas violently interacting with incident laser light, leading to occurrence of great electric field intensity and Raman enhancement in specific nanoscale regions called hot spot [28,29]. CME is mainly a microscopic effect where the molecule Raman cross section (or Raman activity) is enlarged by the formation of loose chemical bond between molecule and surrounding background and consequent action of additional pathways of charge-transfer resonance [29,30].

The above argument can be placed into a more clarified physical model. The signal intensity of enhanced Raman scattering is expressed as

$$I(\omega_R) = \{AI_0(r_0,\omega)\} \times |\alpha(\omega_R,\omega)|^2 \times \left\{\frac{|E(r_0,\omega)|^4}{|E_0(r_0,\omega)|^4}\right\}, \quad (2)$$

where $A$ is a coefficient related in practice with the collection efficiency of the optical system used to collect the Raman signal, $I_0(r_0,\omega)$ is the intensity of incident light, $\alpha(\omega_R,\omega)$ is the Raman polarizability (proportional to Raman activity or cross section), of the detected molecule, and $\frac{|E(r_0,\omega)|^4}{|E_0(r_0,\omega)|^4}$ corresponds to the famous local field enhancement factor. Eq. (2) has implied that one can harness the macroscopic strategy of Raman instrument improvement (first term), the microscopic strategy of molecule Raman polarizability enhancement (second term, CME), and the mesoscopic strategy of local field enhancement (third term, EME) to push up the overall enhancement of Raman detection down to single-molecule (SM) level. Since the space for macroscopic instrument performance improvement is very limited, one should focus to dig into mesoscopic EME and microscopic CME for SM Raman detection.

Previous studies have shown that EME can become large in SERS substrates made from noble metal Au and Ag single nanoparticles with sharp tips, corners, and edges [31,32], or nanoparticle dimers or aggregates with nanoscale gap between them [33-35], or in TERS configuration with a sharp Ag/Au tip sitting above planar Ag/Au substrates [36-38]. So far, the biggest EME occurs in Ag/Au nanoparticle dimers or aggregates, which can reach 10-11 orders of magnitude, still 3-5 orders of magnitude shorter than the SM detection limit. Compared with EME, CME achieves a relatively low enhancement factor. In usual Ag/Au nanoparticle SERS substrates, CME is present together with EFE, but the magnitude is only 10-100. In some carbon nano-porous substrates [39], CME can reach a very high value of $10^6$, but unfortunately, the CME is negligibly small here. It is clear that either EME or CME alone is unable to make SM Raman detection, but synergic action of simultaneous strong EME and CME might be feasible to bring reality the dream of SM Raman detection.

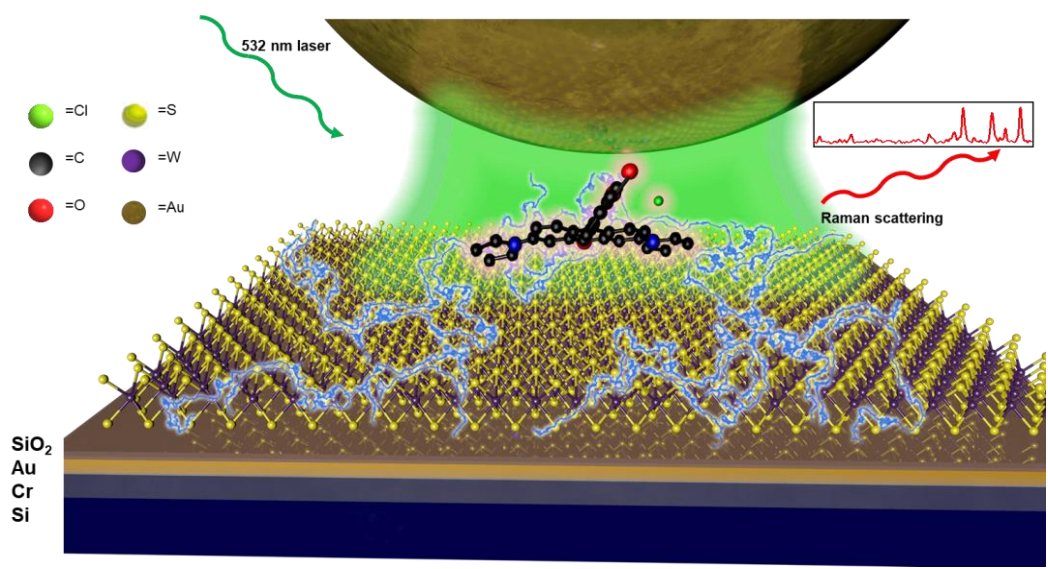

Fig. 1. The golden sphere on the top refers to the attached Au nanoparticle. The green region below is the hot spot of excitation laser. The black (carbon), mazarine (nitrogen) and red (oxygen) ball-and-stick model presents the RhB molecule. The yellow (sulfur) and violet (tungsten) ball-and-stick model shows the mono-layer $WS_2$. The blue flash pattern illustrates the charge transfer between the RhB molecule and mono-layer $WS_2$. The subjacent cuboid shows the silicon slice with several coatings.)

## Results

In this work we harness both EME and CME to reach a very high level simultaneously. We successfully design and implement a 2D material-plasmon nanogap composite nanoscale system and make unambiguous experimental observation of single-molecule Raman spectroscopy of RhB molecule. The RhB molecules are absorbed on two-dimensional monolayer $WS_2$ sheets, within the plasmonic nanogap formed by gold nanoparticle sitting upon 2 nm $SiO_2$ coated gold thin film on a 5 mm × 5mm square silicon wafer as shown in Fig. 1. Raman spectroscopy down to an extremely dilute value of $10^{-18}$ mol/L can be clearly visible, and the statistical enhancement factor could reach 16 orders of magnitude compared with the reference detection sample of silicon plate with a detection limit of $10^{-2}$ mol/L. The electromagnetic enhancement comes from local plasmon resonance induced at the nanogap, which could reach ~10-11 orders of magnitude, while the chemical enhancement comes from monolayer WS2 2D material, which could reach 4-5 orders of magnitudes.

We proceeded a series of tests on Mxenes, black phosphorus and transition-metal dichalcogenides (TMDCs), which were promising to have Raman enhancement. A kind of TMDCs $WS_2$ was proved to be capable of $10^4 \sim 10^5$ Raman magnification for RhB, while the other tested materials did not show enhancement factor higher than $10^2$, so we estimated $WS_2$ as the most promising CME substance for RhB SM Raman probing. RhB molecules were attached on monolayer $WS_2$, and afterwards laid on a 2 nm $SiO_2$ coated gold thin film. Finally, gold nanoparticles were added to form nano-gap with the Au plating (Fig. 2(a)). Optical microscope image shows the tight attachment of $WS_2$ slices on the base (Fig. 2(b)), and scanning electron microscopy (SEM) picture in Fig. 2(c) indicates that the Au nano-particles were distributed barely evenly on the mono-layer $WS_2$ (each with area of one square micrometer). Energy dispersive spectroscopy (EDS) elemental mapping clearly demonstrates that the tungsten element is distributed on the gold film uniformly (Fig. 2(d)), in another word, the monolayer $WS_2$ is uniformly attached to the base[40].

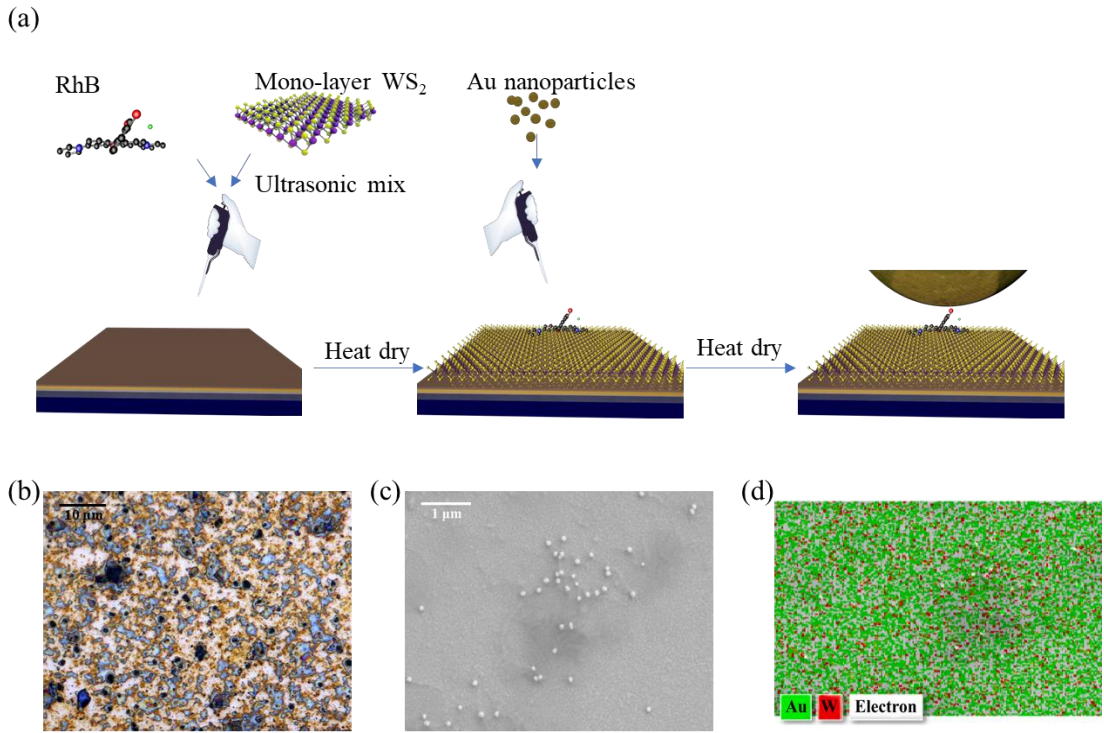

Fig. 2. (a) Preparation progress of the sample. (b) Optical microscope and (c) SEM image of the base surface. (d) EDS elemental mapping of the gold nanoparticle on the base.

Spectra in Fig. 3 illustrates the CME effect of monolayer $WS_2$ for RhB with various concentrations, which indicate the lowest detectable threshold of pure RhB is $10^{-2}$ M in a silicon substrate with neither EME nor CME for Raman scattering. With the aid of the $WS_2$ CME, the detection limit of RhB molecule can reach $10^{-7}$ M, meanwhile the Raman intensity of $10^{-6}$ M RhB on $WS_2$ is of same order of magnitude with $10^{-2}$ M sole RhB on silicon wafer. Thus, the CME factor of monolayer $WS_2$ can be defined as $10^4$ ~$10^5$.

CME for $WS_2$ based SERS is dependent on the electronic structure of the interface between the RhB molecules and the $WS_2$, therefore the molecules layer closest by the substrate is the key crucial to CME, which is generally defined as "first layer effect" [41]. The RhB molecule contains a chloride anion and a cation framework, thus the molecule frameworks can attach tightly to the $WS_2$, which has abundant negative charge on the surface. Moreover, the charge transfer (CT) mechanism tends to appear between the adsorbed RhB molecules and $WS_2$ mono-layer material. On the basis of frontier molecular orbital theory, the electrons with different energies in molecules will be assigned into different molecular orbital energy levels [42], among which the highest occupied molecular orbital (HOMO) and the lowest unoccupied molecular orbital (LUMO) are dominating in the interaction with other matters [43,44]. The calculated molecular orbitals of RhB and energy band of monolayer $WS_2$ indicates that the valence band of $WS_2$ is totally included between the HOMO and LUMO of RhB. As a result, charge carriers in the valence band can provide charges to the HOMO through charge transfer (CT), and enhance LUMO via CT resonance in the meantime (Fig. S1, S2). The extremely high carrier mobility of $WS_2$ (Table S1) further amplifies the effect of CT and CT resonance, so that finally a CME of $10^4$ ~$10^5$ can be achieved [45-48].

Because of the charge transfer between the RhB molecules and $WS_2$, several peaks in the $WS_2$ enhanced Raman spectra of RhB fluctuate or/and shift compared with the normal spectrum, which are declared by grey dashed lines in Fig. 3. Let "X" represents the xanthene ring, "M" represents the methyl, and "D" represents the diethylamino group, respectively. $WS_2$ enhanced Raman spectra of RhB differ from the ordinary spectrum in the following details. The bands of $C_X$–$C_X$–$C_X$ bending vibrations are not shifted to different wavenumbers, while the broad band at 613 cm$^{-1}$ turns to an extremely narrow and strong peak and the band at 632 cm$^{-1}$ decreases dramatically. In the case of the $C_X$–H in-plane bending vibrations, the band at 717 cm$^{-1}$ almost disappears, the band at 1136 cm$^{-1}$ gets weaker and shifts to the lower wavenumbers. In the mean time, the band at 761 cm$^{-1}$ standing for $C_X$–H out-of-plane bend is shifted to higher wavenumbers at 773 cm$^{-1}$, while the $C_X$–$C_X$ stretching vibrations (1193, 1367, 1510, 1560 cm$^{-1}$) are shifted to lower wavenumbers (1182, 1359, 1499, 1542 cm$^{-1}$). Meanwhile the $C_X$–N stretching band appears around 1389 cm$^{-1}$. The above results suggest the spatial orientation reform of the RhB molecule to the $WS_2$ substrates, and CT enhancement takes place on the interface[49]. It can be explained that the "X" is approximately parallel to the surface when RhB is adsorbed on $WS_2$, hence, the $C_X$–$C_X$ stretching bands are highly enhanced other than the $C_X$–N stretching band and the $C_X$–H bending vibrations[50].

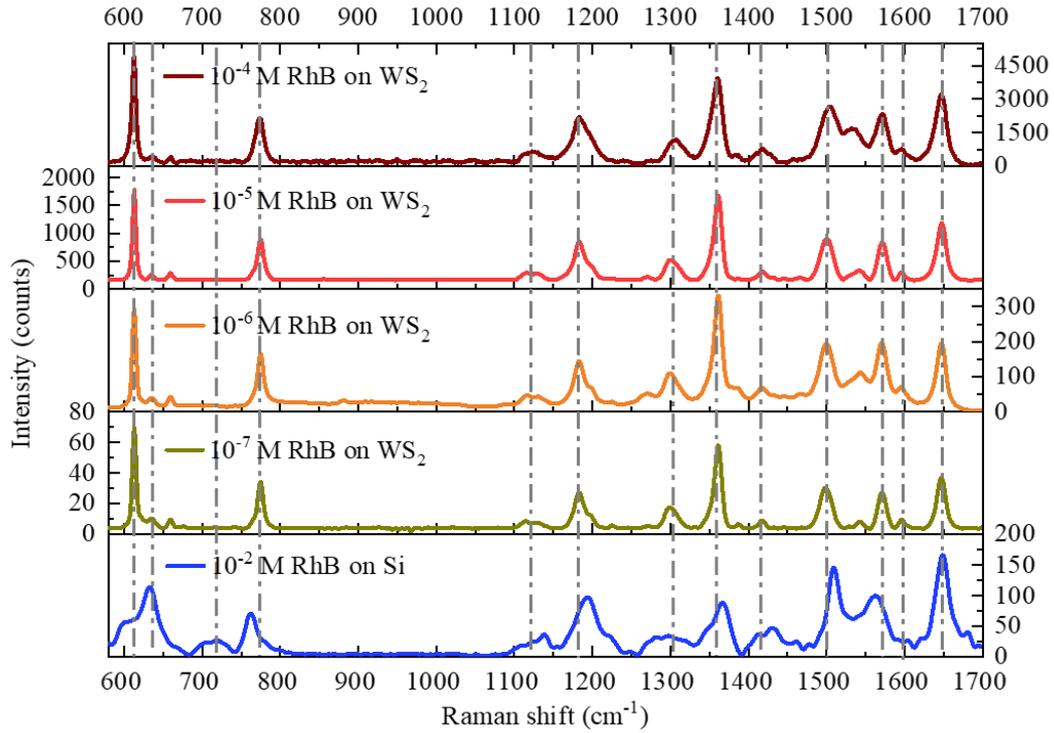

Fig. 3. WS$_2$ enhanced and normal Raman spectra of RhB. Molecular mechanism of WS$_2$ chemical enhancement for RhB.

For a more quantitative comparison, we have carried out three-dimensional finite-difference time-domain (3D-FDTD) simulations to get the electromagnetic enhancement factors with and without the WS$_2$ layer and the RhB molecule (SIII.). The simulation results show that the existence of WS$_2$ layer can enhance the effect of local electromagnetic field, so can the Rhodamine B molecule. Summarly, the maximum electromagnetic enhancement factor can approach to the order of $10^9$ according to the FDTD simulations. However, in practical situations, the maximum EF values have the capacity to reach an even larger orders of magnitude when taking into account more complex sample surface and optical field pattern under experiment conditions. For instance, the roughness and shape anomaly of the gold nanoparticles and plating could lead to a larger EME. Besides, molecular Raman polarizability could become higher due to the quantum confinement from WS$_2$ film and nano-gap upon RhB molecule. Consequently, the overall EME factor can be levitated up to the level of $10^{10}$-$10^{11}$.

The above results reveal that with matched 2D semiconductor materials, the Raman signal of molecule can be amplified $10^5$ times due to the chemical interaction between substances, known as the CME of SERS. Whereas the EME mechanism does not rely on the microscopic-level light-matter interactions, it is more like an environment construction: noble metallic nanoparticles and nanostructures form nanogaps with hot spots, where enormous electric field intensity and Raman enhancement take place in this specific nanoscale regions due to the LSPR [51,52]. Thus, through appropriate combination, the CME and EME can make a synergy to gain single-molecule level Raman detection power. A series of control-experiments was carried out to verify the feasibility.

The experiments were taken on 5 mm ×5 mm chips for each sample. Every chip was added by 30 μL RhB solution with different concentrations. When the concentration reaches down to $10^{-18}$ M, there were only 18 molecules on one chip on average. The distribution of Au nanoparticles in samples with gold nanogap is about one every square micron. The scanning scope for the samples is 40 μm ×60 μm, the sampling step is 1 μm. The diameter of facula area is easily calculated as 1.298 μm. The weakest detected Raman signal of $10^{-6} \sim 10^{-20}$ M RhB samples are shown in Fig. 4(a), which obviously indicates that with the lowest concentration detection limit as $10^{-18}$ M, the 2D $WS_2$ material incorporated gold nanogap SERS substrates have successfully entered the realm of SM Raman detection.

It is noteworthy that under ultra-low concentrations, even when the amount of total molecules is increasing, the Raman signal remains in the SM level. This is because only the molecules which have obtained both EME and CME simultaneously can be detected. Besides, as the above results have shown, the detection limit of CME is just $10^{-7}$ M, and within the incident laser region, there is only one EME hot spot which can contain just an individual RhB molecule [12]. Therefore, when the RhB concentration reaches $10^{-6}$ M, the signal intensity begins to grow and the value is approximately equal to the sum of former-mentioned SM signal and $10^{-6}$ M RhB on $WS_2$ in Fig. 3, as the other exposed molecules initiate to contribute observable Raman signal together with the single molecules sitting inside the hot-spot. A 40 μm ×60 μm range in the center of each sample was scanned, and the achieved Raman mapping pictures of the intensity at 1650 cm$^{-1}$ peak are exhibited in Fig. 4(b). Furthermore, the whole chip of 10$^{-20}$ M sample was completely scanned, and no signal could be detected. When the quantity of molecules rises, the amount of active hot spot also increases to induce more SM signals. Taken together, by the synergy of CME and EME, unambiguous RhB SM Raman detection has been achieved. Under extremely low concentration, RhB molecules were dispersed in the form of single molecule. Because the LSPR in nanogap only occurs in a small region for only a single molecule, Raman signal detected in low concentration samples can only be attributed to SM signals. Meanwhile, the increase of molecule amount will result in more active hot spots to take action and thus more intense Raman signal to be detected. We have compared the Raman mapping picture and the corresponding photomicrograph (Fig. S4(a) as an example), and confirmed that a particle sitting right upon the molecule absorbed WS$_2$ is the essential requirement to achieve SM-SERS signal, in another word, the synergy effect of EME and CME is vital to SM-SERS. To further verify the repeatability of our method, we randomly chose two points on several samples with signal detected. The detected curves are shown in Fig. S4(b). The signal to noise ratio (Fig. S4(c)) of these curves are calculated as the ratio of the mean valve to the standard deviation of them, which is fluctuating between 6 ~8.5.

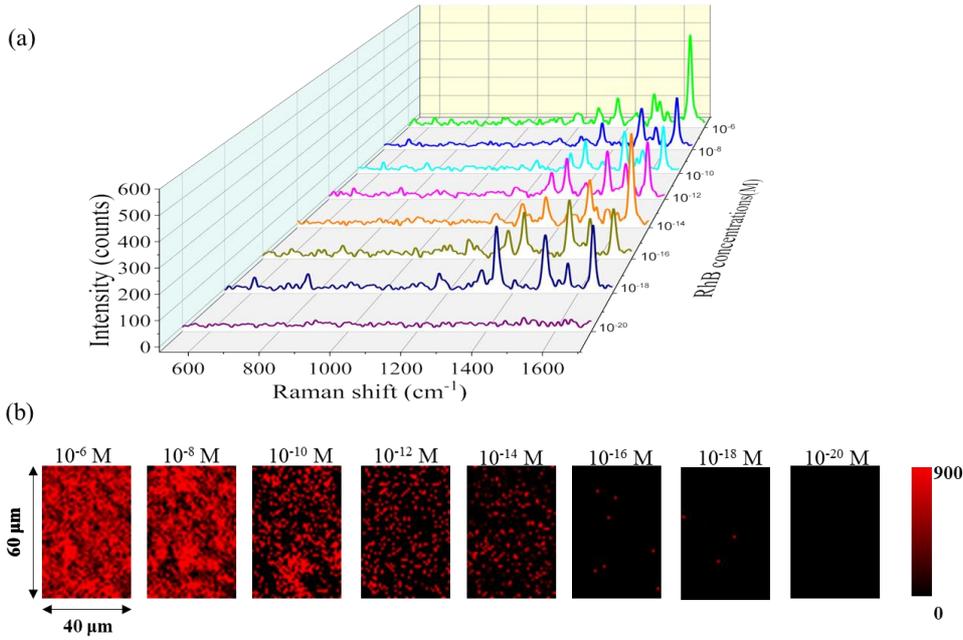

**Fig. 4.** (a) Raman spectra of RhB with variable concentration in SM Raman base. (b) Raman mapping of the intensity at 1650 cm$^{-1}$ peak of RhB at concentrations $10^{-6}$ ~ $10^{-20}$ M.

## Discussion

In conclusion, we have dug deeply into the two major enhancement mechanism for SERS, the EME and CME, explored to maximize the strength of each mechanism, and more importantly, and found out the optimal synergetic action of them, in a hope to fulfill the condition of SM Raman detection. We have found experimentally that 2D material monolayer $WS_2$ offers a very promising CME with enhancement factor up to $10^5$ for absorbed RhB molecules, due to strong CT and CT resonance between the high-mobility $WS_2$ substrate and RhB molecules. Besides, it is well-established that the plasmonic nanogap made from gold nanoparticles sitting on the gold thin film can offer a very promising EME with enhancement factor up to $10^{10}$ for molecules absorbed within the nanogap hot spot under an appropriate gap separation size. We have designed and implemented a hybrid configuration of plasmonic nanogap formed by a gold nanoparticle sitting on the 2D monolayer $WS_2$ and 2 nm $SiO_2$ coated gold thin film, which in principle should enable synergic action of EME and CME with their individual enhancement factors multiplied. Indeed, our experiments have unambiguously shown that Raman spectroscopy for RhB molecules absorbed on the hybrid plasmonic nanogap can still be clearly observed with an extremely low concentration down to $10^{-18}$ mol/L, and has successfully entered the realm of SM Raman spectroscopy. The statistical enhancement factor could reach an unprecedented high value as $\sim 10^{16}$ for this hybrid nano-device when compared with the reference detection sample of silicon plate. Moreover, the designed and implemented SM Raman base with so outstanding power of SM detection is very easy to fabricate. Thus, further researches on different molecules and applications are promising. Such a synergic route of Raman enhancement devices could open up a new frontier of single molecule science, allowing detection, identification, and monitor of single molecules and their spatial-temporal evolution under various internal and external stimuli.

## Method

**Materials**

Rhodamine B was obtained from Shanghai Adamas Reagent Co., Ltd. Mono-layer $WS_2$ dispersion was purchased from Nanjing MKNANO Tech. Co., Ltd. Gold nanoparticle colloid was achieved from Nanjing XFNANO Materials Tech Co., Ltd. The gold, chromium and $SiO_2$ sputtering target were acquired from Beijing Dream Material Technology Co., Ltd. All reagents were of analytical grade and used directly without further purification. Deionized water was produced through a Millipore water purification system (Milli-Q, Millipore) and used throughout the study.

**Instruments and measurements**

The SEM image and EDS elemental mapping were taken by a GemniSEM 500 electron microscope. The HR-TEM images were recorded using a JEOL-2010 electron microscope. The gold, chromium and $SiO_2$ coating on the SM Raman base were sputtered by an MSP-300B magnetron sputtering instrument. The thickness of coatings was measured by a Bruker DektakXT step profiler. All the Raman spectra were gained from a Renishaw inVia Raman spectrometer, laser: 532 nm edge (mode: Confocal), grating: 2400 l/mm (vis), exposure time for each point: 0.5 s, laser power: 2.5 mW, slit opening 20 μm, objective ×50 L.

**Fabrication of the RhB SM Raman base**

20 nm chromium, 300 nm gold and 2 nm $SiO_2$ were sputtered on a 5 mm ×5 mm square silicon slice, successively. 0.2 mL $WS_2$ (0.1 mg/mL) dispersion was mixed with 2.8 mL RhB solution with specific concentration, the mixture was ultrasonic stirred for 30 minutes to make the RhB molecules absorbed adequately by the 2D material. Afterwards, 20 μL of the mixture was dropped uniformly on the coated slice. after 2 hours vacuum drying under 40 °C, 20 μL gold nanoparticle (6 μg/mL) colloid was added on the aforementioned slice. Eventually, via 2 hours vacuum drying beyond 40 °C, the RhB SM Raman base can be created.

## Availability of data and materials

The data that support the findings of this study are available from the corresponding author on request.

## Abbreviations

*CME:* Chemical enhancement

*EME:* Electromagnetic enhancement

*EF:* Enhancement factor

*2D:* Two-dimensional

*RhB:* Rhodamine B

*SM:* Single molecule

*SERS:* Surface-enhanced Raman spectroscopy

*SPP:* Surface plasmon polariton

## Acknowledgements

Not applicable.

## Funding

This work is supported by the National Natural Science Foundation of China (11974119), Science and Technology Project of Guangdong (2020B010190001), Guangdong Innovative and Entrepreneurial Research Team Program (2016ZT06C594), and National Key R&D Program of China (2018YFA 0306200).

## Authors information

**Authors and Affiliations**

School of Physics and Optoelectronics, South China University of Technology, Guangzhou 510640, China

Haiyao Yang, Haoran Mo, Jianzhi Zhang, Lihong Hong & Zhi-Yuan Li

State Key Laboratory of Luminescent Materials and Devices, South China University of Technology, Guangzhou, 510640, China

Zhi-Yuan Li

**Authors' contributions**

Z. Y. Li supervised the project. Z. Y. Li and H. Y. Yang conceived and designed the experiments, H. Y. Yang prepared the samples and performed the experiments, L. H. Hong offered technical assistance, H. R. Mo and J. Z. Zhang offered assistance in theoretical calculations. H. Y. Yang and Z. Y. Li provided the theoretical and experimental analysis, and wrote the manuscript. All authors participated in the discussion of results and reviewed the manuscript.

**Corresponding author**

Correspondence to Zhiyuan Li: phzyli@scut.edu.cn

## Ethics declarations

**Ethics approval and consent to participate**
There is no ethics issue for this paper.

**Consent for publication**
All authors agreed to publish this paper.

**Competing interests**
The authors declare that they have no competing interests.